\documentclass[multphys,vecphys]{svmult}

\usepackage{makeidx}         % allows index generation
\usepackage{graphicx}        % standard LaTeX graphics tool
\usepackage{multicol}        % used for the two-column index
\usepackage[bottom]{footmisc}% places footnotes at page bottom

\makeindex             % used for the subject index
\begin{document}

\title*{UV Spectroscopy of Metal-Poor Massive Stars in the Small Magellanic
Cloud}

\titlerunning{UV Spectroscopy of Massive Stars in the SMC}
% your contribution title if the original one is too long

\author{Daniel J. Lennon\inst{1,2}}

% Use \authorrunning{Surname1 et al.} for an abbreviated version of
% the authors names title if the original one is too long

\institute{Isaac Newton Group of Telescopes
\texttt{djl@ing.iac.es}
\and 
Instituto de Astrofisica de Canarias 
}
%
% Use the package "url.sty" to avoid
% problems with special characters
% used in your e-mail or web address
%
\maketitle

\begin{abstract}

The Hubble Space Telescope has provided the first clear evidence for 
weaker winds of metal-poor massive stars
in the Small Magellanic Cloud, confirming theoretical predictions of the 
metallicity dependence of mass-loss rates and wind terminal velocities.
For lower luminosity O-type stars however, derived mass-loss rates are
orders of magnitude lower than predicted, and are at present unexplained.

\end{abstract}

\section{Introduction}
\label{sec:1}

Massive stars may loose mass throughout their lifetimes either as
steady winds or in episodic outbursts duing a short-lived
Luminous Blue Variable (LBV) phase.  To first order, the 
product of a massive star's mass-loss rate during its core-hydrogen 
or core-helium burning phase times the duration of these evolutionary
phases can amount to a significant fraction of the star's initial
mass.  Similar considerations apply to the shorter-lived LBV phases.
Mass-loss therefore plays a crucial role in the evolution of stars
more massive than about 20 solar masses, uncertainties of
even a factor of two being significant for the star's evolution.
There is general agreement that the steady winds of massive stars
are driven by radiation pressure on millions of metal absorption
lines, these winds therefore depend on metallicity ($Z$).  Calculations
\cite{Puls} indicate that in the range
$0.1Z_{\odot}<Z<3Z_{\odot}$ the mass-loss rate $\dot{M} \sim Z^x$ where 
$x\sim 0.7$, while  the ratio of wind terminal velocity to
escape velocity ($v_{\infty}/v_{esc}$) is rather
insensitive to $Z$, though dependent on effective temperature.
The Small Magellanic Cloud with $Z\sim0.2Z_\odot$
is therefore the ideal
laboratory in which to study the wind properties and evolution of massive
stars at low metallicity.

\section{Weak winds in the Small Magellanic Cloud}
\label{sec:2}

\begin{figure}
\includegraphics[width=6cm]{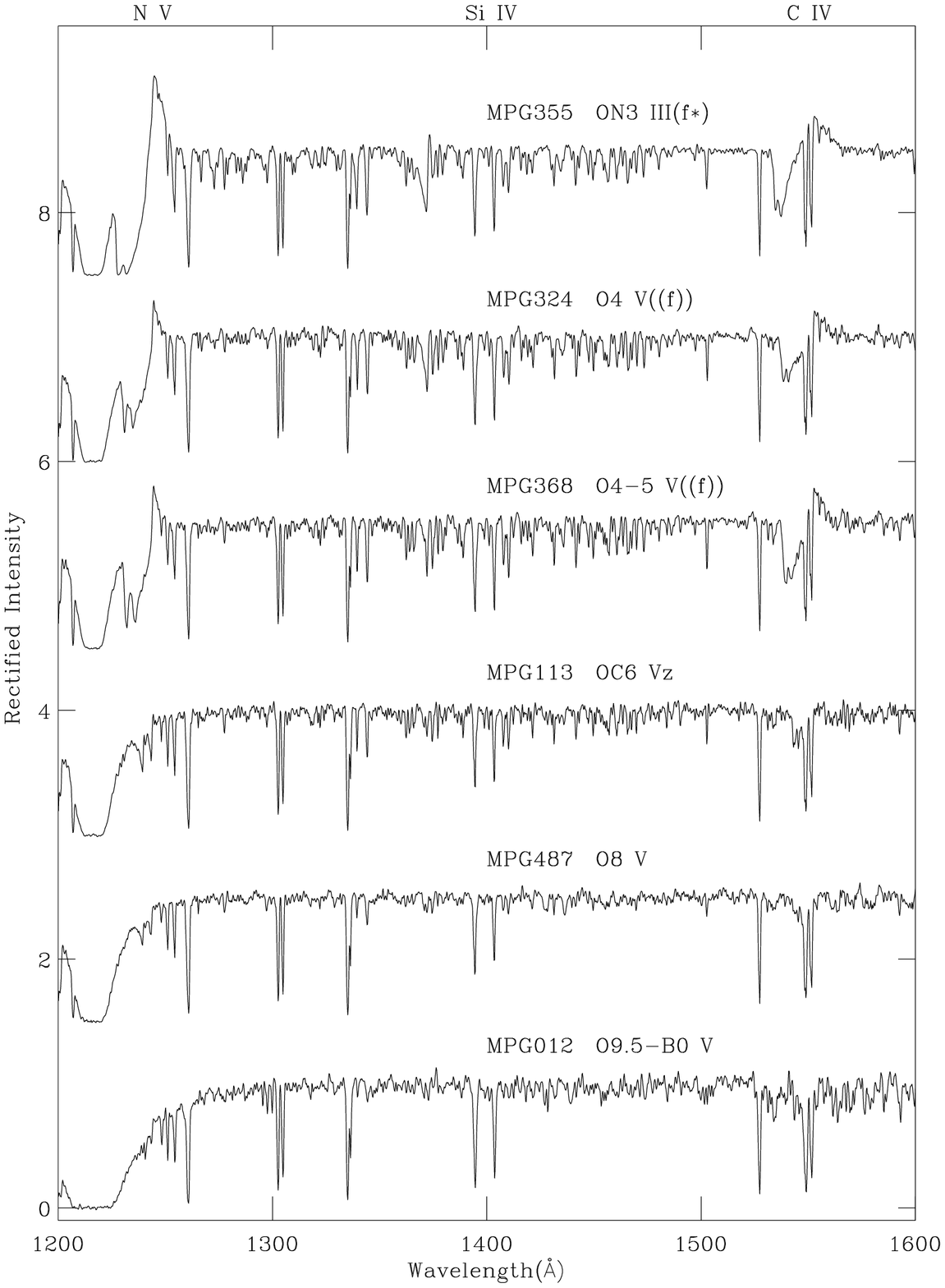}
\hspace{\fill}
\includegraphics[width=6cm]{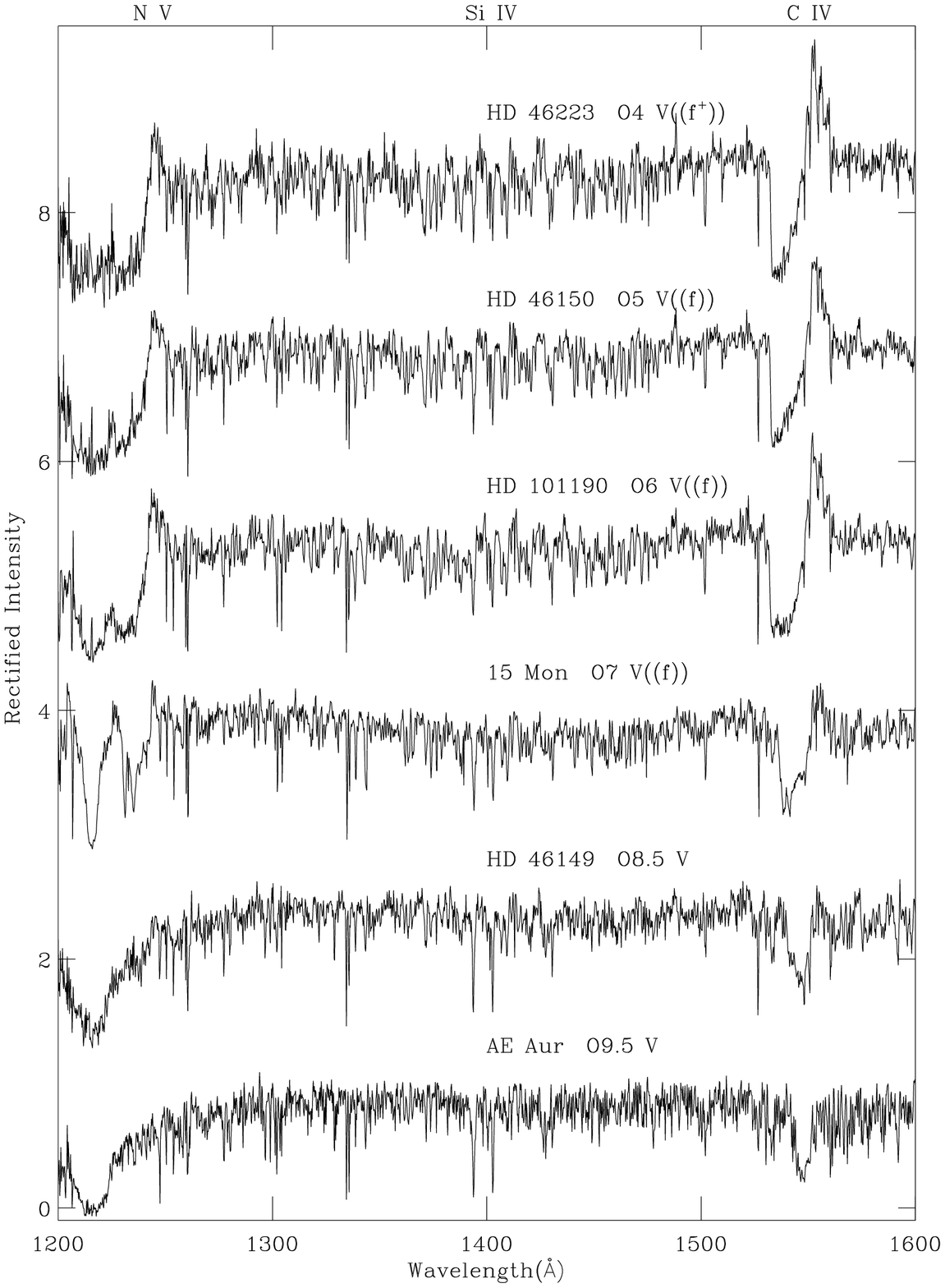}
\caption{Spectra of O-type main-sequence stars. Left panel; SMC
sources observed with HST/STIS, all of which lie in the 
cluster NGC\,346.  
Right panel; Galactic sources 
observed with IUE at high resolution \cite{Wal85}. 
The features
identified at the top are N\,{\sc v} $\lambda\lambda$1239, 1243; 
Si\,{\sc iv} $\lambda\lambda$1394, 1403; 
and C\,{\sc iv} $\lambda\lambda$1548, 1551.
A characteristic of O-type dwarfs in the SMC is the absence of
saturated C\,{\sc iv} P-Cygni profiles, in sharp contrast to
the Galactic stars. In addition wind features are extremely weak
or absent for dwarfs later than O5. The strong N\,{\sc v} feature in 
MPG\,355 reflects the large nitrogen enrichment in this object.}
\label{fig:1}
\end{figure}

\begin{figure}
\includegraphics[width=6cm]{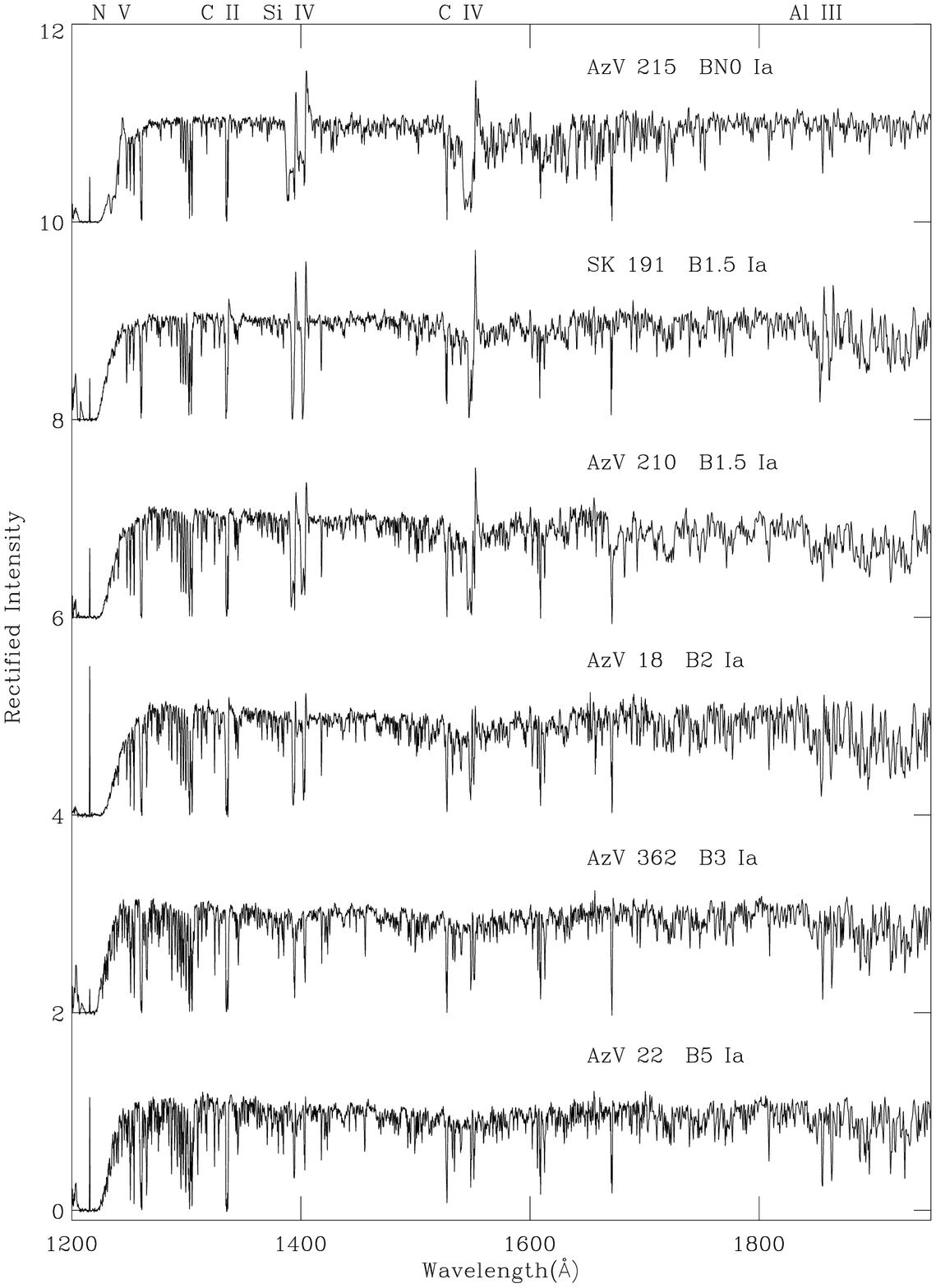}
\hspace{\fill}
\includegraphics[width=6cm]{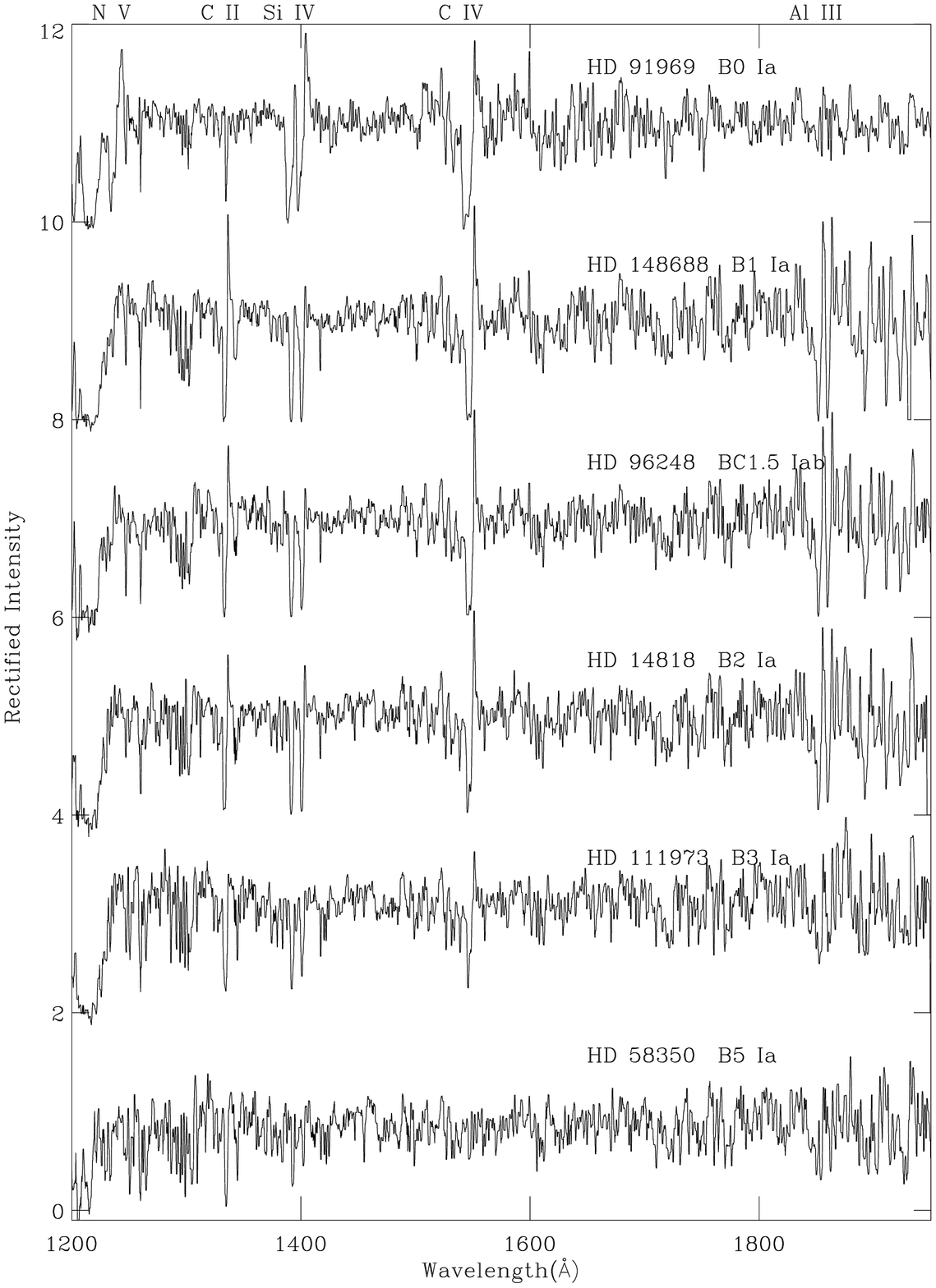}
\caption{Spectra of B-type supergiant stars. Left panel; SMC
sources observed with HST/STIS. Right panel; Galactic sources 
observed with IUE at high resolution \cite{Wal95a}. 
The features
identified at the top are N\,{\sc v} $\lambda\lambda$1239, 1243; 
C\,{\sc ii} $\lambda\lambda$1334, 1336;
Si\,{\sc iv} $\lambda\lambda$1394, 1403; 
C\,{\sc iv} $\lambda\lambda$1548, 1551;
and Al\,{\sc iii} $\lambda\lambda$1855, 1863.
Note the extreme weakness of these features in the SMC stars compared
to their Galactic counterparts.
}
%\leftcaption{This is the caption for the figure on the left}
%\rightcaption{And this is the caption for the figure on the right}
\label{fig:2}
\end{figure}

There were several attempts to confirm theoretical expectations 
using the IUE satellite to observe massive stars in the Magellanic
Clouds, however with mixed reults. The Hubble 
Space Telescope (HST) revolutionized our UV view of massive stars in
the Magellanic Clouds, with the first high s/n UV spectroscopic
campaign (Kudritzki, GOs:2233/4110) yielding startling spectra from the
Faint Object Spectrograph (FOS), whose
morphology confimed the presence of weak winds in the O-type main-sequence
stars in the SMC \cite{Wal95b}. Subsequent quantitative
analyis confirmed that mass-loss rates were indeed lower in the 
SMC sample than in LMC and Galactic counterparts  \cite{Puls96},
and in reasonable agreement with theory. This prompted a more
systematic approach to the SMC, enlarging the previous sample of 9 stars 
with an additonal 30, the use of high resolution with STIS, 
the inclusion of B-type supergiants, and an extension to lower
luminosity late-O dwarfs (Lennon, GOs:7437/9116; \cite{Wal00},
\cite{Evans04a}). The UV spectra
for O-type dwarfs and B-type supergiants are displayed and
discussed in Fig.\ref{fig:1} and Fig.\ref{fig:2} respectively.

%\section{Anomalous weak winds in the SMC}
%\label{sec:3}

\begin{figure}
\centering
% Use the relevant command for your figure-insertion program
% to insert the figure file.
% For example, with the option graphics use
\includegraphics[height=6cm]{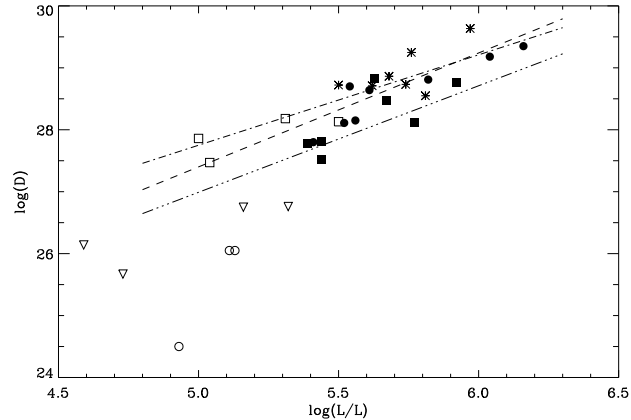}
%
% If not, use
%\picplace{5cm}{2cm} % Give the correct figure height and width in cm
%
\caption{Modified wind-momentum ($D=\dot{M}v_{\infty}\sqrt{R}$) versus luminosity
for SMC OB stars observed with HST (\cite{Bou}, \cite{Hill}, \cite{Tru04}, 
\cite{Tru05}); filled symbols repesent stars with
measured terminal velocities, open symbols are stars with assumed terminal
velocities, circles and squares repesent O- and B-type stars respectively.
Additional OB supergiants from \cite{Evans04c} are marked with asterisks.In all
cases we refer to un-clumped mass-loss rates.
The lines are the empirical fits derived for O-type stars\cite{Mok} 
(dashed line), early B-type supergiants \cite{Tru05} (dash-dot) and
mid B-type supergiants \cite{Tru05} (dash-dot-dot-dot).
The three
O-type stars (open circles) well below the expected relationship,
plus the upper limits (inverted triangles) from 
\cite{Mar} represent the anomalous weak wind class of O-type stars.}
\label{fig:3}       % Give a unique label
\end{figure}

Many of these stars have been analysed to derive mass-loss rates
confirming the expected theoretical trend of mass-loss with metallicity
(\cite{Bou}, \cite{Hill}, \cite{Tru04}, \cite{Tru05}, \cite{Evans04c})
while theoretical
expectations for terminal velocities have also been confirmed \cite{Evans04b}.
However, a small number of stars were found to have mass-loss rates
which are an order of magnitude or more less than theoretical predictions
\cite{Bou}, see Fig.\ref{fig:3}. 
Additional members of this group have been found in the SMC \cite{Mar}, 
in all cases these stars are low luminosity O-type dwarfs
having anomalous weak winds. 
The nature of their winds is not as yet
understood, but possibly indicate an incomplete 
understanding of mass-loss in the weak wind limit, or perhaps arise 
as a result of the lack of accurate diagnostics.

%
%
% BibTeX users please use
% \bibliographystyle{}
% \bibliography{}
%
% Non-BibTeX users please follow the syntax here below:

%%%%%%%%%%%%%%%%%%%%%%%%%%%%%%%%%%%%%%%%%%%%%%%%%%%%%%%%%%%%%%%%%%%%%%  }

%%%%%%%%%%%%%%%%%%%%%%%%%%%%%%%%%%%%%%%%%%%%%%%%%%%%%%%%%%%%%%%%%%%%%%

\printindex
\end{document}